\documentclass[BTech]{iitmdiss}
\usepackage{graphicx} 
\usepackage{t1enc}
\usepackage[driverfallback=dvipdfm]{hyperref}
\title{thesis}
\begin{document}


\title{PRIORITISING INTERACTIVE FLOWS IN DATA CENTER NETWORKS WITH CENTRAL CONTROL}

\author{S Mohana Prasad}

\date{APRIL 2016}
\department{COMPUTER SCIENCE}

\maketitle
\certificate

\vspace*{0.5in}

\noindent This is to certify that the thesis titled {\bf PRIORITISING INTERACTIVE FLOWS IN DATA CENTER NETWORKS WITH CENTRAL CONTROL}, submitted by {\bf S Mohana Prasad}, 
  to the Indian Institute of Technology, Madras, for
the award of the degree of {\bf Bachelor of Technology}, is a bona fide
record of the research work done by him under our supervision.  The
contents of this thesis, in full or in parts, have not been submitted
to any other Institute or University for the award of any degree or
diploma.

\vspace*{1.5in}

\begin{singlespacing}
\hspace*{-0.25in}
\parbox{2.5in}{
\noindent {\bf Krishna Sivalingam} \\
\noindent Research Guide \\ 
\noindent Professor and Head of Department \\
\noindent Dept. of CSE\\
\noindent IIT-Madras, 600 036 \\
} 
\hspace*{1.0in} 
\end{singlespacing}
\vspace*{0.25in}
\noindent Place: Chennai\\
Date: 29th April 2016

\acknowledgements

I would like to start by thanking my parents for their utmost dedication to my education and my sister and relatives for their affection, support and encouragement.

I would like to thank the Indian Institute of Technology, Madras for giving me the opportunity to work
on a project and my guide Prof. Krishna Sivalingam, for allowing me to work with him and his unparalleled help
and guidance in both academic and personal matters. I would also like to thank Prof. Hari Balakrishnan for the wonderful summer internship opportunity at Massachusetts Institute of Technology, USA and for allowing me to extend that work for my undergraduate thesis. My gratitude goes to Jonathan Perry, PhD student at MIT for his guidance and help throughout this project. 

For their help, constructive discussions, support, motivation, personal advice and friendship, my special thanks go to Manoj Kumar Sure, Ranjan Rajagopalan, Krishnan Raghavan, Aravind Srinivas and Sanjay Ganapathy. Thanks are also due to many others in DON lab for their valuable suggestions and questions at various times. 


\abstract

\noindent KEYWORDS: \hspace*{0.5em} \parbox[t]{4.4in}{Datacenter Networks; Fastpass; prioritizing flows; Congestion Control; SDN; TCP; ECN;}

\vspace*{10pt} Data centers are on the rise and scientists are re-thinking and re-designing networks for data centers. The concept of central control which was not effective in the Internet era is now gaining popularity and is used in many data centers due to lower scale of operation (compared to Internet), structured topologies and as the entire network resources is under a single entity's control. With new opportunities, data center networks also pose new problems. Data centers require: high utilization, low median,tail latencies and fairness. In the traditional systems, the bulk traffic generally stalls the interactive flows thereby affecting their flow completion times adversely. In this thesis, we deal with two problems relating to central controller assisted prioritization of interactive flow in data center networks.

Fastpass from the MIT's lab is a state-of-the-art providing the above properties needed by a data center network. But, unfortunately the central arbiter of Fastpass doesn't scale well for more than 256 nodes (or 8 cores). In our test runs, it supports only about 1.5 Terabits/s of network traffic. In this work, we re-design their timeslot allocator of their central arbiter so that it scales linearly till 12 cores and supports about 1024 nodes and 7.1 Terabits/s of network traffic. The design of the allocator is not specific to Fastpass and can be used in any central arbiter that needs a timeslot allocator.

In the second part of the thesis, we deal with the problem of congestion control in a software defined network. We propose a framework, where the controller with its global view of the network actively participates in the congestion control
decisions of the end TCP hosts, by setting the ECN bits of IPV4
packets appropriately. Our framework can be deployed very
easily without any change to the end node TCPs or the SDN
switches. We also show 30x improvement over TCP cubic and
1.7x improvement over RED in flow completion times of interactive traffic for one
implementation of this framework. 



\begin{singlespace}
\tableofcontents
\thispagestyle{empty}


\end{singlespace}







\pagebreak
\clearpage

\pagenumbering{arabic}


\chapter{INTRODUCTION}
\label{chap:intro}

In the Internet, most of the network decisions are taken in a distributed way. For instance, path selection decisions are distributed among the routers and the packet transmission and congestion control decisions are distributed among the end nodes (sometimes supplemented by the routers).

With the rise in the number of data centers, people have started re-thinking and re-designing networks specifically for data centers. The concepts that applies for the Internet need not apply to data centers anymore. The requirements have changed. For instance, data centers are generally controlled and operated by a single entity and is not very chaotic and unstructured as the Internet. More control also means a known topology and better optimizations specific to the scenario. Data centers are also generally owned by rich companies who keep it updated with bleeding edge technology. 

Data center infrastructure and hardware generally supports high link rates. But, data centers have to accommodate multiple users and a mix set of workloads. For instance if we take a Facebook data center, it needs to support a wide range of traffic. There might be large amounts of bulk traffic like server migration, map-reduce job running , etc. At the same time, there might be a user trying to retrieve his photo from the Facebook servers. The interactive traffic from the users are very critical for their operation and hence they should ensure that this interactive traffic is serviced with high priority. The main problems faced by data center operators is that they have less control over the network. A better fine gained control can help ensure lower end to end latencies.

One main advantage with data centers is that they are smaller in scale when compared to the Internet. We can exploit the concept of central control were we have dedicated servers that take network decisions. In recent times, especially in data center networks central control is becoming famous. With the emergence of Software Defined Networking, central control has gained more importance and is being deployed in many data center networks. 

There are some properties that needs to be provided by an ideal data center network: high utilization, low median,tail latencies and fairness. High utilization is important so that we make use of the resources available. Low tail latencies guarantee lower flow completion times for interactive flows. Consider a 2 MB photo that needs to be loaded from a Facebook server. The median latency might be small, but the photo will not be rendered to the user till the last packet arrives. Tail latencies become important for reducing the flow completion times and to better the quality of service. Fairness again translates to better quality of service where we allocate network resources so that the interactive traffic is not stalled by bulk flows.

In this thesis, we deal with two problems relating to Data center Networks. Both these problems involve a central controller and are aimed at prioritizing interactive flows. In the first problem, we take an existing system (Fastpass) and propose and evaluate ways to scale it up for larger data centers. In the second problem, we propose our own congestion control framework for a software defined network and show results on a specific implementation that prioritizes interactive flows.

FastPass [6] is a state of the art data center architecture that takes an extreme approach for providing the properties that we discussed earlier. In Fastpass each
sender delegates control to a central arbiter who ensures zero
queuing in intermediate routers, by deciding when an end node
should send a packet into the system and also selects a path for every packet. An end node before sending a packet into the system, sends a request to the central arbiter and the arbiter allocates a timeslot and a path for this packet. The central arbiter ensures the "zero queuing" property at the routers.

With the zero queuing property, queue sizes won't raise and fall, variance in latencies will be small, tail latencies will be under control. Besides eliminating persistent congestion, there also won't be any packet drops due to buffer overflows at the routers and hence no packet retransmissions. 

There are other advantages that come with the central arbiter. Now that there is zero queuing and the arbiter has the freedom to allocate timeslots for packets, I can prioritize flows as I want giving us extreme control over the network. We can give higher priority to interactive traffic and stall bulk traffic if need be.

Even though Fastpass has shown impressive results like 240x reduction in queue length, 5200x better standard deviation in per sender throughput, 15x better median ping time during their test run at Facebook data centers.

Fastpass is plagued by the problem of scaling up. Fastpass works for 256 node network and the central arbiter was able to manage only about 1.5 Terabits/s of network traffic on 8 cores during our test runs of FastPass (code made available by the authors). It doesn't scale beyond 8 cores and it also doesn't scale linearly from 1 to 8 cores.

Fastpass is based on two fast algorithms, a timeslot allocation algorithm and a path selection algorithm. The present design of the timeslot allocator restricts the scalability. In the first part of the thesis we deal with multi-core algorithm design and software architectures for the problem of timeslot allocation that scales linearly for large number of cores. This timeslot allocator that we design is not specific to Fastpass and the same technique should come handy to many other timeslot allocators and central arbiters. We start off by analysing the existing architecture and propose some changes and evaluate the results. We then design, implement and analyse two architectures and discuss its advantages and shortcomings. We show linear scaling up to 12 cores where we support 7.1 Terabits/s of network traffic. We also show that we can support  more than 1.5 Terabits/s of network traffic on just 3 cores as opposed to 8 cores as used by the current timeslot allocator of fastpass. A detailed introduction and explanation of all the methods can be found in the third chapter.

In the second part of the thesis, we deal with data center networks in the Software defined network setting. We find that interactive traffic often get penalized because of the distributed congestion control algorithms. We propose our own congestion control framework that can ensure the properties of data center networks that we discussed earlier.

Congestion control deals with the problem of, when
an end node should send a packet into the system to enable
efficient and fair sharing of the network resources. Most of the
congestion control algorithms are distributed, where the end
nodes take their decision on congestion control based on the
limited information that they get, either packet drops or delays.
Some end-to-end schemes are improved by explicit router participation, such as Explicit Congestion Notification (ECN). TCP has
been built to work on a broad range of network conditions and
it tries to adapt its congestion window without much knowledge
about the underlying network and traffic characteristic. With the
emergence of SDN and central control, we can leverage the global
view of the network and its traffic to make better congestion
control decisions. In this work, we explore the advantages of
bringing in global information into distributed congestion control.
We propose a framework, where the controller with its global
view of the network actively participates in the congestion control
decisions of the end TCP hosts, by setting the ECN bits of IPV4
packets appropriately. Our framework can be deployed very
easily without any change to the end node TCPs or the SDN
switches. We also show 30x improvement over TCP cubic and
1.7x improvement over RED in flow completion times for one
implementation of this framework.

A more detailed introduction about both the works can be found in the respective chapters.

This thesis has been organized in the following way. The Background and Related Work chapter talks about the Fastpass system and also goes into relevant works in congestion control. We have the third chapter dedicated for giving a detailed explanation about our various techniques and architectures for scaling up Fastpass to larger data centers. Fourth chapter deals with our ECN based Congestion Control technique for a SDN network. The final chapter concludes and summarizes the thesis work.


\chapter{BACKGROUND AND RELATED WORK}
This chapter has been organized into two parts, the first part explains "Fastpass: A Centralized "Zero-Queue" Data center Network" that forms the basis for our work on "Scaling up Fastpass for large data centers". The second part of this chapter focuses on congestion control which lays the background work for our fourth chapter "ECN based congestion control for a software defined network".

\section{Fastpass}

Fastpass [6] is a data center network architecture where we have fine-grained control over transmission times and network paths through a central arbiter. Fastpass aims at high utilization and zero queuing at the routers. It provides this "zero queuing" property by, each sender delegating control to a centralized arbiter which decides when a packet has to be sent into the network and what path it has to take. Fast pass takes a rather extreme approach has complete control over the network. 

To understand Fastpass better, lets consider a road traffic control system. Tradition road traffic systems have signals at junction (analogous to routers) to moderate traffic. The traffic waiting at he signals are analogous to queue length at routers. Instead consider a hypothetical traffic control system, where there is a central controller (central arbiter) and every time a car wants to go from location $X$ to location $Y$, it sends a request to the central controller and the controller assigns the car a timeslot at which it needs to depart and also assigns the path it has to take. The central controller takes care of the zero queuing property at the junction and hence the cars can just zoom past without waiting at any of the signals. This is exactly what Fastpass does for a data center network where the road system is the network, junctions/signals are routers and cars are the data packets. The packets gets queued in the sender's queues rather than getting queued at the routers. We will soon discuss the advantages of doing this.

In traditional network, whenever there are bursts of packets, these are absorbed by queues at the routers. The queue sizes may rise and fall, so does the end to end latencies. The median latencies can be low, but the tail latencies are generally found to be high. With the zero queuing property, queue sizes won't raise and fall, variance in latencies will be small, tail latencies will be under control. Besides eliminating persistent congestion, there also won't be any packet drops due to buffer overflows at the routers and hence no packet retransmissions. 

There are other advantages that come with the central arbiter. Now that there is zero queuing and the arbiter has the freedom to allocate timeslots for packets, I can prioritize flows as I want giving us extreme control over the network. We can give higher priority to interactive traffic and stall bulk traffic if need be.

Fastpass is based on two fast algorithms: the timeslot allocator and the path assignment algorithm. Whenever a sender applications calls $send()$ on a socket, the OS sends a request to the central arbiter with the source, destination and the number of bytes to be sent. The arbiter first runs the timeslot allocation algorithm and allocates a set of timeslots at which it has to send its packets. The arbiter then runs the path assignment algorithm for each packet and sends the timeslot and path information back to the requesting node. The arbiter has the complete information about all the packets that are to be sent in the network, and hence can ensure "zero queuing". 

This paper uses Fastpass Control Protocol(FCP) for all communications between the nodes and the central arbiter. FCP is a reliable protocol with ACK and some aggregations to ensure that the control packets don't flood the network. Moreover, these FCP packets are set as high priority packets in the intermediate routers and they suffer very less delays.

Fastpass requires some hardware support in the end host NICs  with protocol support in the operating system. Fastpass doesn't need any modification to the routers. The central arbiter needs to process the timeslot and path for every packet sent in the network. As a result, efficient and scalable algorithms become essential at the arbiter. In this thesis we discuss the timeslot allocator (the bottleneck in the arbiter) of Fastpass and we re-design it so that it is more than 2x efficient and scales for large number of cores.

Fastpass was deployed and was tested in a section of Facebook's data center. The results of the paper show that Fastpass achieves 240x reduction in queue length at the routers, 5200x reduction in standard deviation of per flow throughput (means fairer allocation), 15x better median ping time, 2.5x reduction in TCP retransmissions.

In our run of the arbiter (code made available by the authors), it achieves about 1.5 Terabits/s on 8 cores. Current Fastpass doesn't scale up. Fastpass works for 256 node network and it doesn't scale beyond 8 cores It also doesn't scale linearly from 1 to 8 cores.

\newpage
\section{Congestion Control}

Congestion management is a fundamental problem in computer networks, as it enables us to achieve cost-effective
and efficient sharing of network resources across various
end nodes. Managing the shared resources is a vital and
a difficult problem to solve. Congestion control addresses
the fundamental question: When should an end node send
each of the packets of data so that it doesn't throttle the
system. There are many senders sending at the same time
and network may experience different delays, which makes
it a hard problem to solve. Over the past few decades a
plethora of congestion control algorithms have come up, which
are generally heuristics developed based on some specific
assumptions of the network. We would see some congestion control techniques that are relevant to our work.

\subsection {Router-based Congestion Control}

The DECbit mechanism [12] is one of the first works
that used an explicit congestion control protocol to signal the end
nodes about congestion at a router. An router, when it is likely to
experience congestion, reacts by marking a bit at the packet header
for some of the packets. This bit was then sent by the receiver to the
sender for taking appropriate action.

The Random Early Detection (RED) mechanism [10] attempts to
maintain an average queue length at the routers.  Using threshold
values for time-averaged queue lengths, a router drops packets with a
probability that increases with the size of the queue. This packet
drop results in TCP sources reducing their congestion window and hence
transmission rate.

In the Explicit Congestion Notification (ECN) mechanism [3],
packets will be marked with Congestion Encountered (CE) bits instead
of being dropped. The end-node TCP protocol is modified so that the
receiver echoes the CE bits to the sender. When the sender receives
such TCP segments, it reacts by reducing the congestion window.  TCP's
performance can be increased significantly using RED/ECN by setting
the parameters with caution. 

In all these approaches, the decision to drop or mark packets during
congestion is taken in a distributed manner by each router.  In an
SDN-enabled network, it is possible to take advantage of the
(logically) centralized control plane to set the bits based on the
global view.  Our work in chapter 4 presents an attempt to explore the advantages
of pushing this decision to the SDN controller rather than doing it at
the routers.

\subsection{DCTCP}
DCTCP [7] is a variant of TCP optimized for data center
networks. DCTCP requires both, changing the end node TCP
stack and modifying the switch ECN settings to achieve
its objective. DCTCP uses Explicit Congestion Notification
(ECN) at the switches and runs its variant of TCP at the end
nodes. The ECN marks received at the receiver is a single bit
stream, DCTCP tries to provide the sender with a multi-bit
feedback about congestion.

DCTCP tries to address two main problems encountered
in data center networks, 1) interaction of interactive and
bulk TCP flows 2) TCP incast. DCTCP differs from ECN
standard in two ways. DCTCP uses instant queue length at
the switches to trigger CE markings, in place of the average
queue length used by ECN standard. In DCTCP, the receiver
echos the ECE for every packet rather than for a window
of packets. This extra information enables DCTCP sender to
cut its congestion window in proportion to the percentage of
CE marking. If all ACKs for a window of data are marked
with ECE, DCTCP reduces the congestion size to half. The
reduction of congestion window is smoothed out by using
an exponential filter over the ratio of marked packets at the
sender.

The idea of using one CE echo for every packet of the
window allows DCTCP to get multibit feedback about the
congestion in the network. DECbit reduces its window by half when the ratio of marked packets is above a set threshold,
whereas DCTCP cuts the congestion window by the value of
half the fraction of marked packets in the previous window.
This smoothens the reduction of congestion window and
enables DCTCP achieve low queue length still ensuring high
throughput.

Other methods have been developed that proposes switch
modification only schemes, and does not require any changes
in the end nodes. This paper [8] shows that similar performance can be achieved without changing the end nodes TCP
algorithms, but by marking the ECN bits in a clever way at
the routers.

In chapter four, we attempt to solve this problem of congestion
control in the SDN domain, requiring no changes to the end
nodes and with "dumb" SDN switches used today. This makes
the deployment of our framework straightforward.

\subsection{OpenTCP}

OpenTCP [2] is a congestion control framework proposed for
SDN based networks.  The OpenTCP software at the controller collects
information about the underlying network such as link utilization
values. OpenTCP aims at reducing the Flow Completion Times (FCTs) by
updating initial congestion window and retransmission timeout interval
for TCP flows. OpenTCP sends Congestion Update Epistles (CUE) packets
to the end nodes consisting of information about the suggested changes
to their TCP state. OpenTCP's kernel module running in the TCP stack
of the end node updates their TCP variant's state using the
information in the CUE packets. Thus, OpenTCP requires the end node
TCP protocol to be changed.

The approach proposed in our work is SDN-based and using the ECN
mechanism, but does not require any changes to the end nodes' TCP
protocol. This is advantageous since changing end nodes may not be
possible if all the nodes are not under the same administrator,
especially in multi-tenant networks and data centers providing cloud
services.  Our solution uses the ECN which is supported by many popular
running TCP variants. 

In our method, the controller after analysing the current
state of the network, takes congestion control decisions. The
controller explicitly sends flow rules to the switches, asking
them to mark certain packets. The intimation of congestion to
the end nodes are done using ECN bits. Our method keeps
the existing SDN architecture and the end nodes pristine and
exploits the flow rules of SDN to perform congestion control. Our system can be
very quickly integrated into existing data centers, by just
enabling ECN at all the end nodes followed by running our
congestion control module at the controller.

\newpage


\chapter{SCALING UP FASTPASS FOR LARGE DATA CENTERS}
\label{chap:fastpass}

\section{Abstract}
Fastpass is a centralized data center network architecture build in MIT's lab. Fastpass is plagued by the problem of scaling up. Fastpass works for 256 node network and the central arbiter was able to manage only about 1.5 Terabits/s of network traffic on 8 cores during our test runs of FastPass. The timeslot allocator of Fastpass doesn't scale beyond 8 cores and it also doesn't scale linearly from 1 to 8 cores. We start off by analysing the existing architecture and propose some changes and evaluate the results. We then re-design their timeslot allocator of the central arbiter so that it scales linearly till 12 cores and supports about 1024 nodes and 7.1 Terabits/s of network traffic. We also show that we can support  more than 1.5 Terabits/s of network traffic on just 3 cores as opposed to 8 cores as used by the current timeslot allocator of fastpass. The design of the allocator is not specific to Fastpass and can be used in any central arbiter that needs a timeslot allocator. As the timeslot allocator's core computes maximal matching in a bipartite graph, this work can also easily extended to other applications requiring a scalable multi-core implementation of maximal matching in a bipartite graph problem.

\newpage
\section{Introduction}

Fastpass is a centralized data center network architecture that aims at "zero queuing" at the routers. Fastpass is based on central control where each sender delegates control to a centralized arbiter. And this arbiter decides when a packet has to be sent into the network and what path it has to take. Fast pass provides a fine grained control over the network and its resources. 

Fastpass is plagued by the problem of scaling up. Fastpass works for 256 node network and the central arbiter was able to manage only about 1.5 Terabits/s of network traffic on 8 cores during our test runs of FastPass. The timeslot allocator of Fastpass doesn't scale beyond 8 cores and it also doesn't scale linearly from 1 to 8 cores. We start off by analysing the existing architecture, propose some changes and evaluate the results. We then re-design their timeslot allocator of the central arbiter so that it scales linearly till 12 cores and supports about 1024 nodes and 7.1 Terabits/s of network traffic. We also show that we can support  more than 1.5 Terabits/s of network traffic on just 3 cores as opposed to 8 cores as used by the current timeslot allocator of fastpass. The design of the allocator is not specific to Fastpass and can be used in any central arbiter that needs a timeslot allocator. As the timeslot allocator's core computes maximal matching in a bipartite graph, this work can also easily extended to other applications requiring a scalable multi-core implementation of maximal matching in a bipartite graph problem.

Fastpass's timeslot allocation algorithm finds a matching of end points (a list of sender-receiver pairs) that can send packets during each timeslot. The demands for some links of the network for a timeslot can exceed the maximum capacity of that link. The job of the allocator is to ensure that traffic issued in this timeslot does not exceed the capacity of any link. If we can ensure this, then the "zero queuing" property at the routers can be maintained.

For simplicity in understanding lets assume all the end point links have the same capacity (say 10 Gbps for all the calculations in this paper). If one of the end points had a 20Gbps capacity, we can treat it as two 10 Gbps link and everything explained in this chapter will hold good. Data center networks generally have known. fixed topology. They are organized as tiers. Core routers are at the top level, which connects aggregate switches and aggregate switches connect top-of-the-rack switches which in turn connects servers of a rack. 

Fastpass requires the tiers to be rearrangeably non blocking (RNB) [1] and data center networks in general are RNB. RNB property of a network means that if any traffic that satisfies the input and output end link capacities of the network, then that traffic can be routed through the network without any queuing. The RNB property gives us two great advantages. First, the timseslot allocation becomes simpler as we just need to satisfy the end link constraints (input and output). Second, we can do the timeslot allocation and path computation separately. As long as the allocated matchings satisfy the end link capacity constraints, the path selection algorithm is guaranteed to find a path for the traffic. Data center networks in general satisfy the rearrangeably non blocking property.

If the end link capacities are 10 Gbps and lets consider 1500 byte timeslots, then every timeslot is 1.2 micro seconds. So, our algorithm needs to be really fast to be able to make hundreds of allocations in 1.2 micro seconds. Parts of the code was implemented in assembly and SIMD to achieve such speeds. Longer run times were also achieved with parallelism where multiple cores process demands in parallel. In the following sections we will see multi core program design and implementation to enable super fast and critical time slot allocations.

If we decide to find a set of allocations for a timeslot with the maximum number of pairs possible, then we cannot achieve such high computing speeds. Instead we settle for heuristics. Instead of finding maximum matchings , we find maximal matchings. Fastpass's arbiter processes demands in some order and greedily allocates source-destination pairs if possible (does not violate link capacities constraints). When the arbiter is done processing all the demands, we get a maximal matching where no more demands can be allocated. 

\begin{figure}[htp]
    \centering
    \includegraphics[scale=0.25]{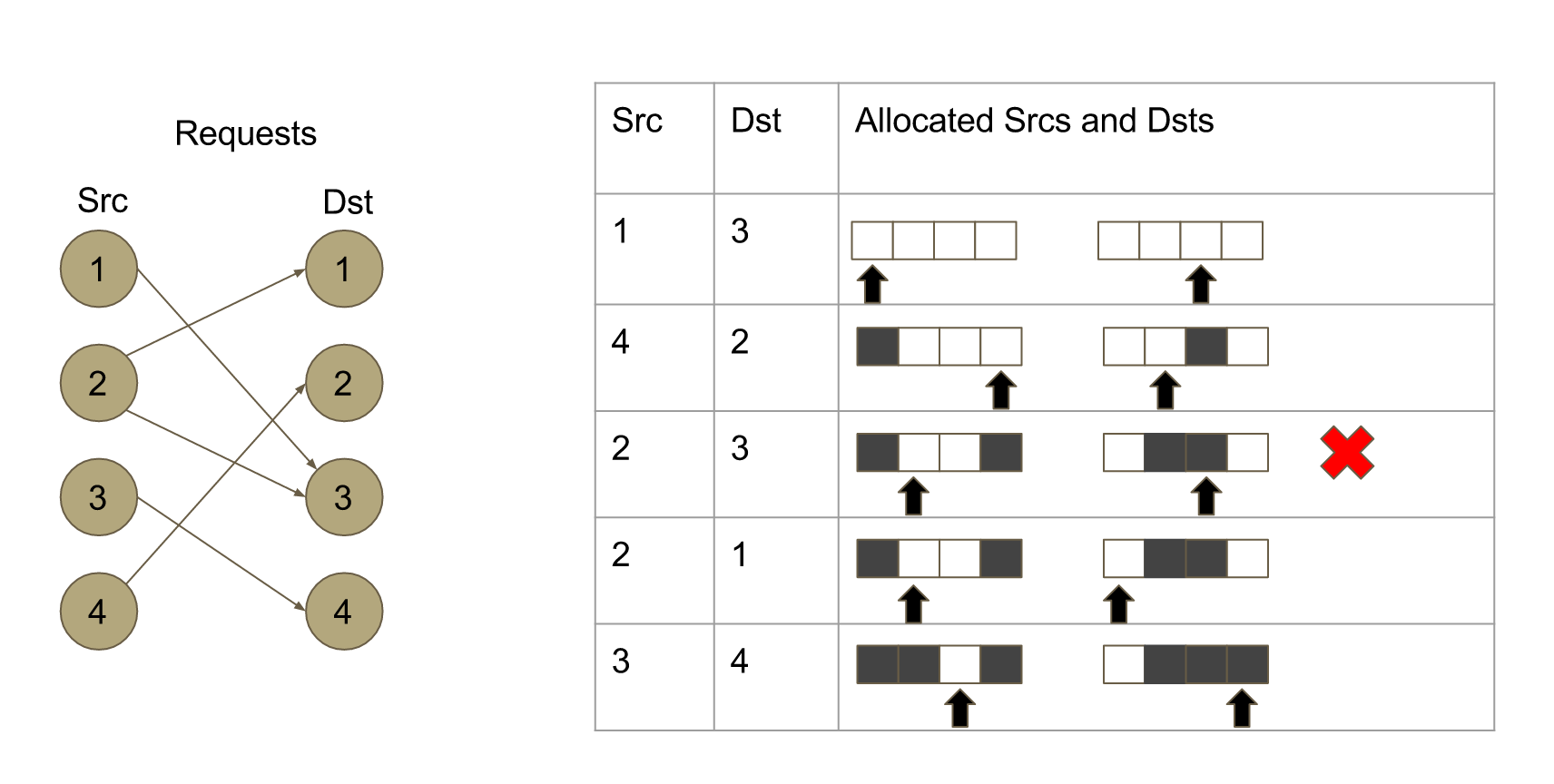}
    \caption{Timeslot Allocation: Greedy heuristic for maximal matching in bipartite graph}
    \label{allocation}
\end{figure}

As discussed earlier, we assume the network is rearrangeably non blocking. Which means, if the traffic satisfies the capacities of the end link (links connecting to the end host servers) then the traffic can always be routed through the network without queuing. While we greedily process requests and try to allocate them, we just need to check for the end link capacity constraints to ensure zero queuing. 

Consider 4 nodes in the system as shown in Figure ~\ref{allocation}. Lets call them nodes 1, 2, 3 and 4. Lets say the capacities of all the links from these end nodes are 10 Gbps. If the demands of the form (source, destination) are processed in this order (1,3), (4,2), (2,3), (2,1) , (3,4), then the first and second demand will be allocated as both the source and destination are available. The third demands cannot be allocated as destination 3 is already allocated. The last two demands can also be allocated. This gives a maximal matching. This can be easily implemented by maintaining a state with two arrays of bits for senders and receivers at each timeslot such that we have a bit for each source and each destination possible. These state bits can be used to maintain the end link capacity constraints. The unallocated demand is considered for allocation in the next timeslot. This is the greedy heuristic algorithm for maximal matching in a bipartite graph problem and this problem has huge applications in many different areas. In this work we would design efficient multi-core implementations for this algorithm.

\newpage
\section{Existing Pipelined Allocator}

In this section we will discuss the Pipelined Allocator Architecture that was used in the Fastpass paper. A detailed understanding of this section is required to appreciate the subsequent sections. This section also contains many details that were not mentioned in the Fastpass paper or the manuals.

Here the CPU cores is logically structured as a pipeline. The core at the head of the pipeline alone reads new demands from a queue called $QHead$. In this section all, core to core communications have been implemented using Intel's DPDK queues. $QHead$ is nothing but a single consumer single producer DPDK queue. The requests from the sender nodes are sent to the arbiter, an exclusive core called the $CommCore$ receives these demands and pushes them into $QHead$ for the $AllocCores$ (the set of cores that we will discuss next) to process the demands.

An incoming demand to the $AllocCores$ is of the following form: (Source, Destination, Number of Packets, Priority Value). We will first see how the cores are organized and then see what happens within an $AllocCore$.

%
%
%
%
%
%
%
%

\begin{figure*}[h]
\centering
\includegraphics[scale=0.25]{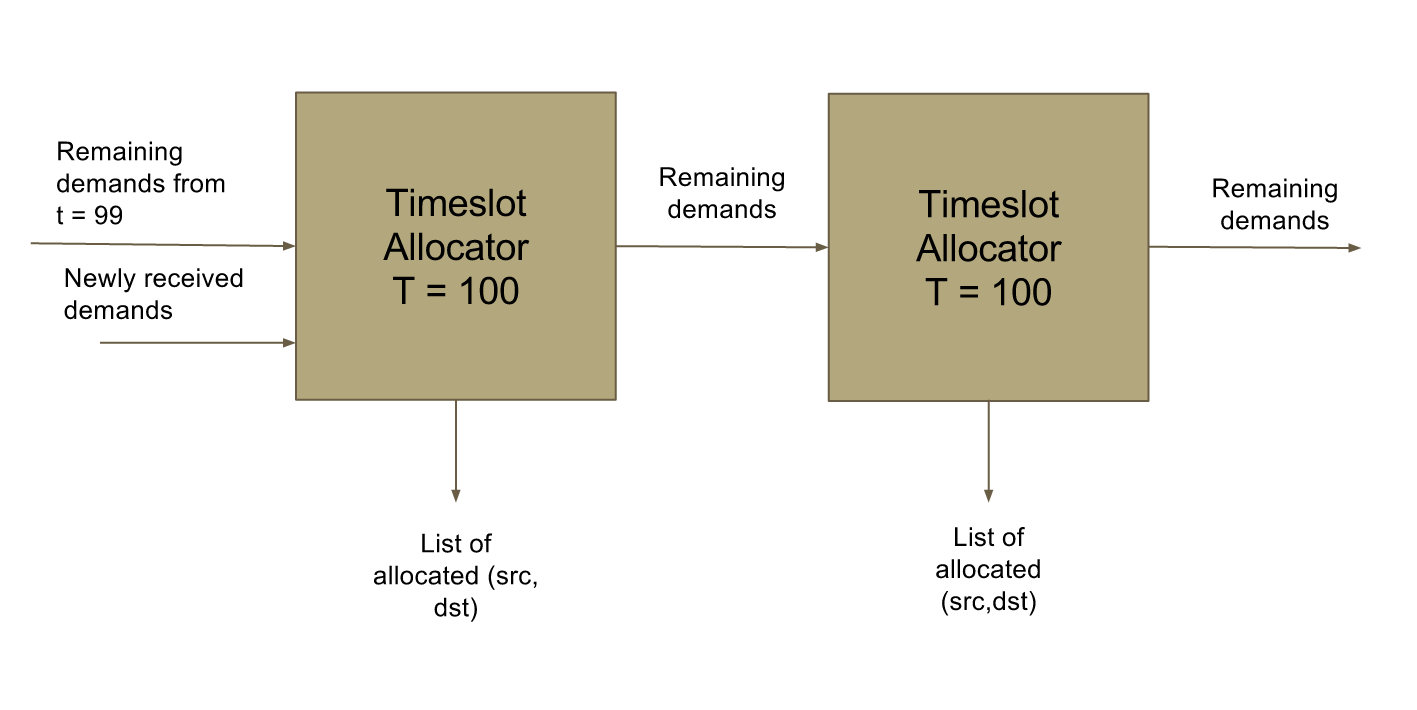}
\caption{Pipelined Timeslot Allocation: The requests unallocated by allocator for time (t-1) goes to allocator of time t }
\label{timeslot}
\end{figure*}

The $AllocCores$ are logically arranged as a pipeline, where the core at the head of the pipeline alone reads new incoming demands from the $QHead$. Each of the alloc cores processes the demands that come into them, tries to allocate as much traffic as possible and it passes on the unallocated demands to the next $AllocCore$ in the pipeline. Each of the $AllocCores$ are processing for a specific timeslot. As shown in ~\ref{timeslot}, the alloc core working on the 100th timeslot gets its input from the core working on the 99th timeslot and this core sends the unallocated demands to the $AllocCore$ processing for 101st timeslot. The head core alone reads input demands from the $QHead$. Assume a system with 8 cores, where the head core is processing for 100th timeslot and the last core is processing for 107th timeslot. Now once the allocations for 100th timeslot is over, the core at the head of the pipeline, logically moves to the end of the pipeline and starts processing for 108th timeslot (gets input demands from alloc core working on 107th timeslot) and the alloc core working on 101th timeslot becomes the new head core and reads new inputs from $QHead$. Such a design makes sure that each core is at the head of the pipeline for 1.2 microseconds (duration of 1 timeslot) but the allocations for 8 timeslots are happening in parallel. A core gets to works on a timeslot for $8*1.2$ microseconds.

On a high level, each of the $AllocCores$ get some input demands, they sort the demands based on priority, then try to allocate these demands using the array of bits as a check for contention. They then pass on the unallocated demands to the next core. Finally when its time (when they are done coming to the head of the pipeline from the tail), they send out the list of admitted demands (the once that are allocated) so that the respective sender nodes can be notified.

%
%
%
%
%
%
%
%

\begin{figure*}[h]
\centering
\includegraphics[scale=0.5]{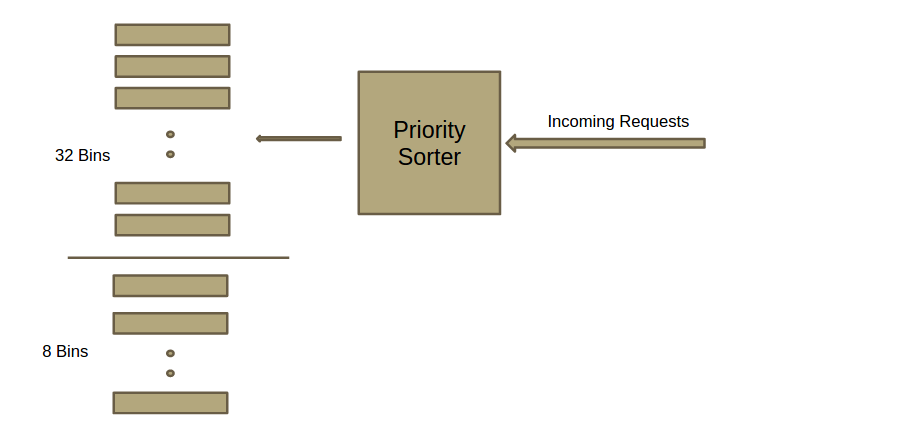}
\caption{Priority Sorting in Alloc Core: Internal bins and the sorter inside the alloc core}
\label{sorter}
\end{figure*}

The priority of a demand is very important to ensure fairness. For max min fairness the core orders the demands by the last timeslot that was allocated to it. The sorting can be a bit tricky. Lets say two contending demands (1,2,100,0) and (3,2, 2, 0) come into the system. The first demands asks for 100 packets to be sent from node 1 to node 2 with priority 0 (last timeslot at which this source destination pair was allocated. Default value of zero)  and the second demand asks for 2 packets to be sent from mode 3 to node 2. If we don't consider priorities and not sort these demands at each of the alloc cores, what will happen is that, the first demand will get allocated at the first core for timeslot 1, then the remaining unallocated demands (1,2, 99,0) and (3,2,2,0) moves to the second core and the first demand gets allocated again. It gets allocated in all the timeslots untill the 100th timeslot. The second demand will just get stalled and will get allocated only in the 101th and 102nd timeslot. So, bulk traffic tend to stall interactive traffic. To avoid such a scenario, whenever we allocate a demand, we set its priority to the timeslot number to which it was last allocated. At every $AllocCore$ we sort these demands based on least recently used policy. this makes sure that we alternatively allocate contending traffic thereby ensuring max-min fairness.

To reduce the overhead of communication between cores, each of the $AllocCores$ allocate for 8 timeslots in one shot and not one timeslot per $AllocCore$. So, a 2 core system will be allocating for 16 timeslots in parallel. We would refer to this number of timeslots each alloc core tries to allocate in one shot as the $BatchSize$. The array of bits that we discussed earlier is now stored as a $bitmap$ for each sender and receiver in the network for each timeslot. For instance, node 1 as source will now have 8 bits called the its 
$bitmap$. A "1" in the bitmap means that node is not scheduled to communicate in that timeslot and "0" otherwise.  To find the first available timeslot for a given packet, the allocator computes the bitwise AND
of the source and destination bitmaps, and then uses the "find first
set" operation (the bsf instruction on x86).

We now take a peek into a single $AllocCore$. We do a coarse grain sorting in the allocator. Lets consider a alloc core working on a $BatchSize$ of 8. We have 32 priority bins as shown in ~\ref{sorter} and we have extra $BatchSize$ number of bins. Initially as the demands come into the system, the sorter does a coarse grain sorting and puts the demands into the 32 bins based on least recently used first priority. A new demand coming later could possible go into one of the higher priority bins, and hence we can't start processing the lower priority bins early on. So, we have a $AllowedMask$ which initially is set to just the top most priority bin and we process demands in the bins that are allowed by the allowed mask alone. As time progresses and the core moves up the pipeline, we slowly relax the $AllowedMask$ to allow more and more of the low priority bins to be processed. By this process we ensure that high priority demands aer processed first. Lets say two demands (1,2,100,0) and (3,2,2,0) comes in and the first demand gets allocated for the first timeslot. Now the demand request becomes (1,2, 99,1). We have seen that the cores process in a batch size of 8. So, we then transfer this demand request to the top most extra 8 bin that we have. We then will process the (3,2,2,0) demand. This demand will be allocated for the second timeslot and then we move the remaining request (3,2,1,2) to the second extra bin. As time progresses, I will start to process the extra bins, starting with (1,2,99,1) demand (as it was put into a high priority bin) and then go on to (3,2,1,2). By such a procedure in place, we can achieve the necessary sorting ability with easy and very less computation.

An adversary end node can send two demands of 50 packets each instead of a single 100 demand request. Or, It could have initially needed just 50 packets, but before these 50 are actually allocated, it might need some more packets to be sent. In such scenarios, duplicate entries for the same source-destination pair can exist in the system. This will affect fairness as each of the 2 requests will be treated independently. To avoid this, there is a $Backlog$ structure in place at the $CommCore$ which accrues these multiple requests and makes sure there is just one copy of a source destination pair request in the system at any point in time. 

\newpage

\section{Analysis of the Pipelined Allocator and Improvements}

\subsection{Test Environment and Initial Results}
To test for scalability and throughput of the arbiter, requests are generated by a synthetic stress-test-core rather than
received from a comm-core. The workload has Poisson arrivals and the senders and receivers are chosen uniformly at random from 256 nodes (for this section, we would later do for 1024 nodes), and number of packets requested is chosen as a gaussian with mean at 10 MTUs. We can vary the mean inter-arrival time (mean\_t) to produce different network loads.

To find out the maximum load the arbiter can take, we run a stress Test with these stress test cores.

The automated test decreases the mean\_t by a constant factor as long as the timeslot allocator is able to approximately match the demand. when the allocator fails, it increases the mean\_t to the last successful value, decreases the constant factor, and repeats. This test makes sure that the Queues do not blow up. The stress test tries to maintain a maximum bound on the  difference between the demand and the number allocated.

This stress test gradually increases and decreases the mean\_t and converges to the maximum load the arbiter can take. An experiment refers to a single run of the stress test for 60 seconds (if not mentioned otherwise).

All the experiments were conducted on MIT's multi core system called Ben. It is 80 core machine with 8 NUMA nodes of 10 cores each. The CPU specification includes: Intel Xeon E7-8870 (06\_2Fh) 2.4GHz, 8 sockets * 10 cores * 2 hardware threads, 30MB shared L3, 256KB L2, 32KB L1 and RAM: 256GB (32x8GB), 1333MHz

Stress test on the existing pipeline gives the following results. This will be the baseline for all our analysis. For 8 cores with a batch size of 8, we get the maximum throughput of 1.27 Tbps. We notice that with 16 cores we perform worse then with 8 cores. Ben has 8 NUMA nodes with 10 cores each. When going to 16 cores, we place cores across NUMA and the communication through QPI could add latency.

\begin{table}[h]
\renewcommand{\arraystretch}{1.3}
\caption{Stress test results on existing pipelined architecture in Terabits/second (Baseline)}
\label{res_pipe}
\centering
\begin{tabular}{|c|c|c|c|c|}
\hline
Batch Size & 2 Cores & 4 Cores & 8 Cores  & 16 Cores \\
\hline
4 &0.477 &0.909 &0.945 &0.782 \\
\hline
8 &0.540 &1.001 &1.273&1.003 \\
\hline
16 &0.592 &1.014 &1.005&0.769 \\

\hline
\end{tabular}
\end{table}






\subsection{Pipeline Analysis}

Scalability may be limited as the cores are doing redundant work. A demand request that gets actually accepted in the 64th timeslot has to by default pass through all the previous 7 cores in the pipeline (assuming batch size 8). Each of these 7 cores attempts to allocate them but fails, and they pass on to the subsequent cores. Finally it gets allocated by the 8th core. This redundancy  may reduce latency. We can defined the optimal scheduler as a magic scheduler that exactly knows (beforehand) at which core a particular demand will get accepted in the pipeline and accordingly schedule that demand to that particular core. An optimal scheduler can prevent any redundant computation.

To test out the above hypothesis we implemented separate log cores to log important variables of the $AllocCore$ and to get a better understanding of what happens inside the alloc core. We study how the cores behave as they progress through the pipeline.

%
%
%
%
%
%
%
%

\begin{figure}[h]
\centering
\includegraphics[scale=0.25]{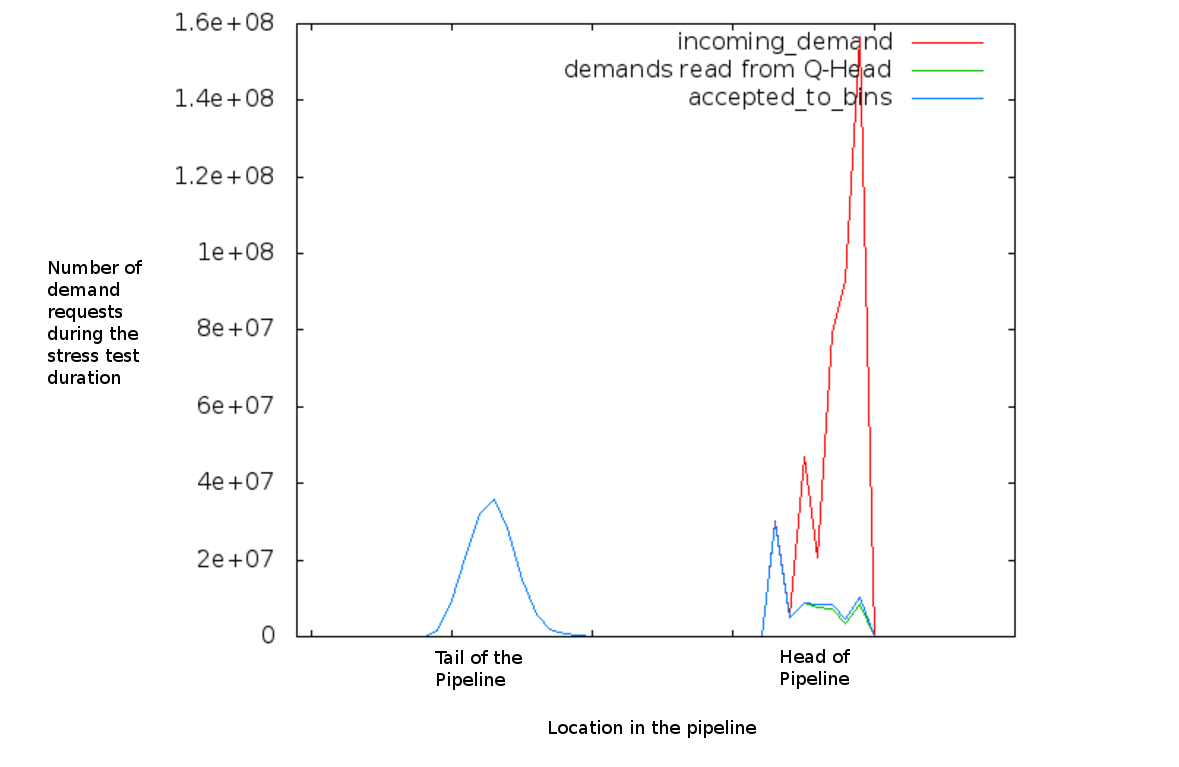}
\caption{Distribution of incoming demands as a function of location in the pipeline.}
\label{demand}
\end{figure}

Figure ~\ref{demand} shows the number of demands coming into the core and number of demands actually accepted/processed by the core as a function of the core's position in the pipeline (for batch size of 8 and 8 cores, a core takes 64 timeslots to reach the head of the pipeline from the tail). This graph shows that the requests coming into the system is in itself bursty. Alloc cores get huge number of requests when they are at the head of the pipeline and generally don't get many requests when they are at the middle of the pipeline. This graph exposes the bursty nature of the requests coming to each cores. 

%
%
%
%
%
%
%
%

\begin{figure}[h]
\centering
\includegraphics[scale=0.25]{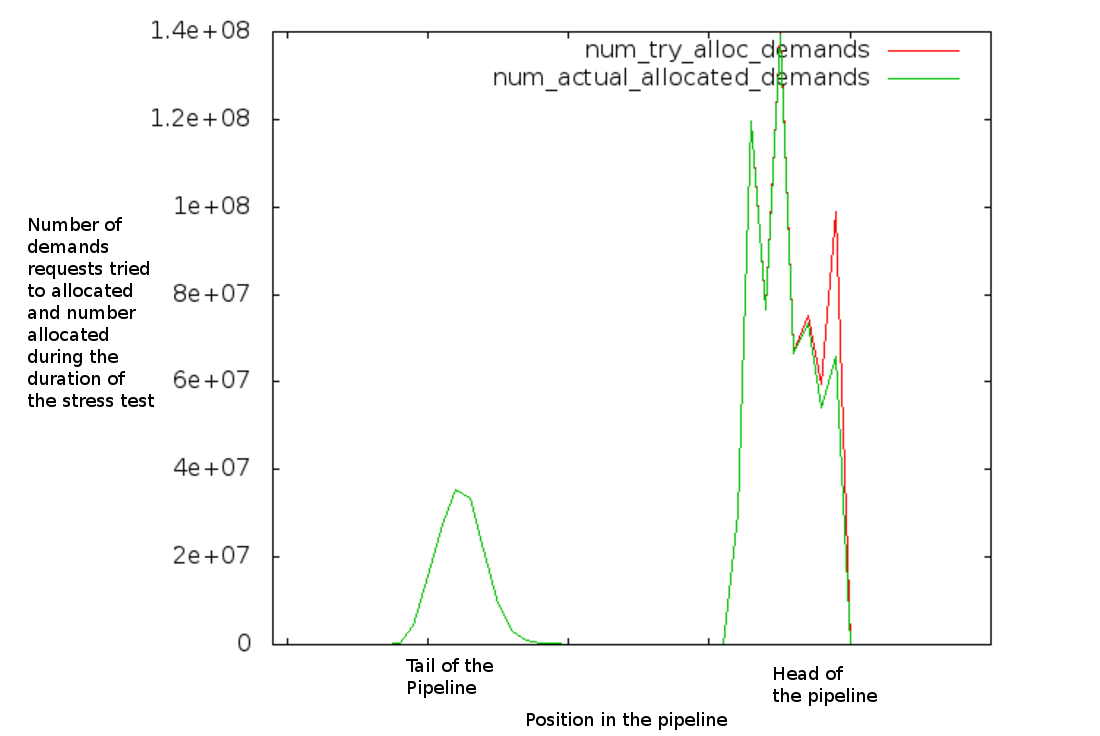}
\caption{Distribution of number of requests tried to allocated and number allocated as a function of location in the pipeline.}
\label{allocated}
\end{figure}

Figure ~\ref{allocated} plots the number of allocations a core tries to make and number of allocations that it is actually able to make as a function of its position in the pipeline. This graph completely disproves our hypothesis. It shows that there isn't much contention among the traffic and hence the redundancy is not the reason for poor scalability. This graph also shows that the cores are idle when they are in the middle of the pipeline and are doing useful work when they are at the head or the tail of the pipeline. The alloc cores are under utilized in the present architecture and are idle most of the times. An ideal plot would be a rectangle where it is utilized to the maximum extent throughout.

It is possible that the demands that come in first are the ones that go into the low priority bins and would not be processed till the end. As a quick check to find out if the allowed mask that we discussed earlier is causing the demands to queue up till the end (and not get processed beforehand), we set up an experiment without the allowed mask feature. Results didn't change much for with and without allowed mask case. This showed that allowed mask wasn't the real bottleneck.

The demands coming in is in itself bursty and not uniform and hence the cores are not getting utilized uniformly throughout. There are a lot of demands coming in during some intervals and there are none during some other times. Presently the cores read new demands (coming from Q\_Head), just before they have to issue admitted bins (when they are in the head of the pipeline). We made the cores read from the queue head and process them, when they are most idle (when they are at the middle of the pipeline). This would also give the cores much more time to process the new incoming demands from Q-Head. We thought such a change would change the distribution of how the demands get passed around the cores. This also means that we add a extra latency for every packet. Even the highest priority packet coming in at this point, cannot be assigned for the immediate timeslot but for some timeslot in future (as it would be sent to the middle core which is processing for a timeslot in future). Such a change gave about 10\% better throughput results in our experiments. 

A detailed analysis revealed that, the extra sorting bins that we discussed in the earlier section was causing the issue. Lets say a request (1,2,100,0) arrives at a core (Batch size 8). Once we allocate it for the first time slot, we push the remaining request (1,2,99,1) to the topmost extra bin. Later in time, when the allowed mask allows to process that that extra bin, we process and allocate it for the 2nd timeslot and push the remaining request (1,2,98,2) to the second extra bin. Similarly we keep pushing the remaining requests to further and further low priority bins within the same core. Finally only when the core is finished processing in the head of the pipeline, we push all these requests out ot the next core. This was the basic flaw with the implementation. The demands were not freely flowing between the cores and were getting accumulated in the head core. As a result the head core always has a lot of load while the other cores don't get anything.

To correct this, we proposed the Batch Processing Mode. This mode ensures fairness on a coarser granularity (granularity of Batch Size).In this mode, once a demand comes into the system, we try to allocate it for the first, second, till the eight timeslot all at one go. we try to allocate as many timeslots as we can within this core (at most batch size) and just send the remaining to the next core. Here for instance, if the demands (1,2,100,0) and (3,2,2,0) come in. At the first core, I read the first demand and will allocate all the eight timeslot slots and just push the remaining request (1,2,92,8) to the next core. Now the first core processes the demand (3,2,2,0) which it cannot allocate and hence just passes on the demand to the next core. Now the second core would have sorted and would read (3,2,2,0) first and would allocate the 9th and the 10th timeslots for it. and will allocate the rest 6 timeslots to (1,2,92,8) and push the remaining demand to the next core. In such a system, demands come into the core, get sorted, get processed, and just goes out of the core as remaining request or as a empty done request. This ensures that the demands don't get accumulated at any core.

\subsection{Results with Batch Processing Mode}
The batch processing mode gave about 1.5x better throughput. We were able to reach throughputs of 1.9Tbps as opposed to our baseline 1.2Tbps. A detailed table can be seen here and we will analyze them a bit deeper in the next section.

\begin{table}[!t]
\renewcommand{\arraystretch}{1.3}
\caption{Stress test results of Batch Processing Mode in Terabits/second}
\label{res_batch}
\centering
\begin{tabular}{|c|c|c|c|}
\hline
Batch Size & 4 Cores & 8 Cores  & 16 Cores \\
\hline
4 &1.197 &1.912 &1.062 \\
\hline
8 &1.483 &1.024 &0.894 \\
\hline
16 &0.764 &0.754&0.692 \\

\hline
\end{tabular}
\end{table}




\subsection{Pipelined Allocator Inferences}

If there are not many contentions in the incoming demands, most of the requests will get issued in the first few cores and may not reach the cores at the tail of the pipeline. The cores at the end will be doing very less work. So, long pipelines are unhealthy. and they also have the overhead of memory allocation/access from different sockets. Moving demands to distant cores are costlier.

If the pipeline length is too short, say 2 cores and say batch size of 4. Say a demand (1,2,10) comes in. The head core allocates 4 timeslots, then it goes to the next cores, say it also allocates 4 timeslots. Now the remaining backlog request of (1,2,2) comes back to the original core and should wait till it finishes processing the earlier timeslot. If we have had a third core, that could have processed this backlog in parallel. There are also lots of moving around of demands which degrades performance. 

The maximum throughput has to increase with number of cores and then decrease after a point. This is found to be true in the results we got with the batch processing mode.  

We typically would expect the length of the pipeline to be the maximum number of cores till which the demands may be alive. This is further reinforced by the fact that for larger batch size the peak occurs at smaller number of cores.
We would also expect that if I increase the batch size, the peak should occur earlier. This phenomenon is observed in table ~\ref{res_batch}.

Lets say we determined the optimal pipeline length. Let's consider a simple model without interior bins. Let us say demands come in, cores process it and send them out. If we take a snapshot of the pipeline at any point in time, the input to the second core comes from the first core and so on. So, the second core cannot work faster than the first core. The second core will be utilized lesser than the first, Third core lesser than the second and so on. 
 So, strictly the utilization of the 1st core $>$ utilization of the 2nd core $>$ utilization of the third core $\ldots$ at this instant of time.

This is an inherent constraint in the pipeline architecture which limits the utilization of the cores.
\newpage
\section{Parallel Allocation Architecture}
In the previous section, we discussed the Pipeline Inequality: At any point in time, if I take a snapshot of the pipeline, Input to the ith core comes from the (i-1)th core. So, throughput of ith core is less than (i-1)th core and we found this to be a constraint in our pipelined architecture.

Lets consider an alternative architecture where the pipeline does not exist anymore and so does the constraint. Let consider a architecture where multiple alloc cores simultaneously allocate demands for the same timeslot. Lets say we have 8 cores and now, all these 8 cores are working on the same timeslot. After completing allocation for this timeslot, they all move to the next timeslot. In the pipelined architecture, one core processes one timeslot for $8*1.2$ microseconds, instead here, 8 cores process in parallel for one timeslot for 1.2 microseconds.

All the alloc cores exist separately and they have a shared memory bitmap. Now, they no longer have the constraint of the utilization of the second core being less than that of the first. Moreover if you consider a demand, in the earlier pipeline, it goes to a core (head core) gets sorted, tries to get allocated,  moves to the queue, goes to next core, gets sorted, tries to get allocated, moves to next queue and so on. But, in our new system, it goes to a core, gets sorted, tries to get allocated, gets sorted again,tried to get allocated, etc. A demand always stays in the same core and doesn't move around.

The path each demand takes is shorter, the utilization of the cores are not restricted, so this new system should perform better provided we can do the reconciliation fast enough (may add latency) and we are able to manage giving input to the cores parallely. 

Even though we have a shared bit map, a race condition can arise across cores which may lead to contending demands being allocated. To cancel such wrongful allocations, we have a reconciliation core which checks for zero contention. We say two demands are contending if they have the same source or the same destination.  

Earlier we were allocating different timeslots parallely, but now, we intent to allocate different demands parallely for the same timeslot and then correct our mistakes (if any).

%
%
%
%
%
%
%
%

\begin{figure}[h]
\centering
\includegraphics[scale=0.25]{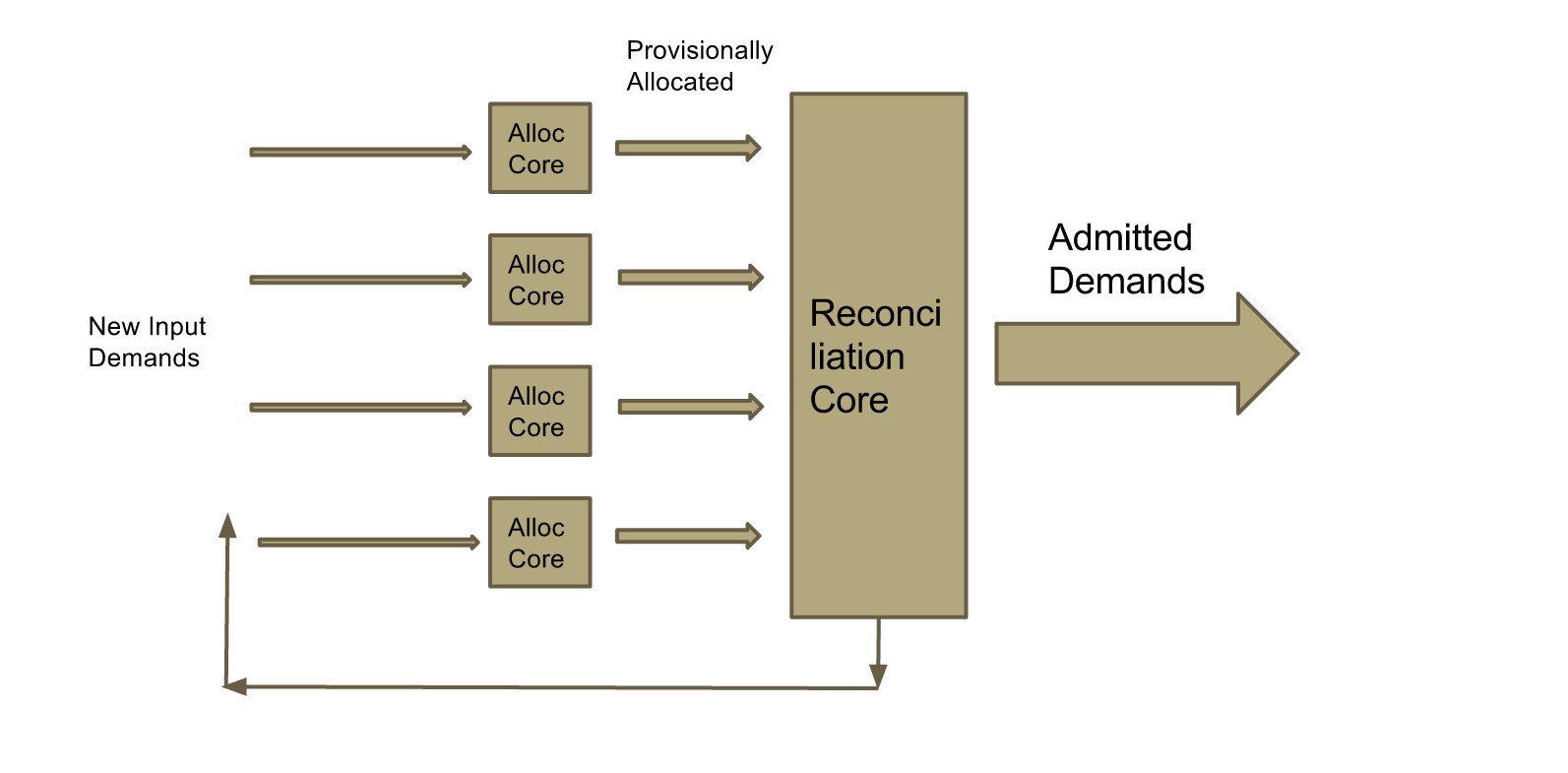}
\caption{Parallel Allocation Architecture}
\label{parallel}
\end{figure}

The figure ~\ref{parallel} shows the architecture diagram. 
\subsection{Implementation Details and Results}

We implement the alloc cores with priority sorter and 32 internal bins. A stress core sends demands to these alloc cores through a multi-producer multi-consumer DPDK queue. The alloc cores read new input from these queues all the time and try to process them. At the end of every timeslot, they send out the list of allocated demands as a admitted bin. They sort and keep the remaining demands for processing in the next timeslot. It was implemented efficiently by keeping two copies of the 32 internal bins and swapping them every timeslot to reduce copying overheads. The following results are for a non shared memory setting, where each core works on its own bitmaps.

\begin{table}[h]
\renewcommand{\arraystretch}{1.3}
\caption{Stress test results of initial implementation of Parallel Architecture in Terabits/second}
\label{res_parallel}
\centering
\begin{tabular}{|c|c|c|c|}
\hline
Batch Size & 1 Cores & 2 Cores  & 4 Cores \\
\hline
8 &0.492 &0.886 &1.614 \\
\hline
16 &0.545 &1.013 &1.890 \\

\hline
\end{tabular}
\end{table}




It scaled linearly till 4 cores. It was not perfectly linear due to cache contention/coherency delays associated with shared memory. Going for the parallel architecture gives 1.5x better throughput than the pipelined architecture by using only half as much the cores.

Instead of multi producer multi consumer queues, we placed multiple single producer single consumer queues. One queue each between the Stress test core and one of the alloc cores. The alloc cores were made to read the inputs in a round robin fashion. The results were similar but slightly better possibly due to lesser overhead in using single producer single consumer queues. The system achieved about 1.95 Tbps and 1.75 Tbps for batch size 16 and 8 for 4 cores respectively.



Even though this architecture performed far better than the pipelined architecture it wasn't able to scale well beyond 4 cores. A thorough analysis with Intel Vtunes profiler exposed that the DPDK queues were costly. A core wasn't able to read/write to more than 4 queues in one time slot. For 8 and 16 cores this architecture broke down.

The shared memory setting did not work well and an detailed analysis for 8 cores with Intel Vtunes profiler showed that 60\% of the cycles are spent in accessing the shared memory due to L1 miss.

This architecture can be used till 4 cores and can give 1.5x better throughput with just half the number of cores required by our baseline.

\newpage
\section {Random Shuffle Architecture}

Random shuffle architecture is based on two main improvements. The parallel architecture wasn't able to scale beyond 4 cores due to the overheads in inter-core communications. In this section we design and implement an efficient distributor data structure for fast inter-core communications. Secondly, we make the job of each core simple and hence fast.

Lets just consider a single alloc core allocating demands for a single timeslot. All the allocator cores that we discussed till now perform complex functions. The allocator cores read/dequeue demand requests, then they sort it into the appropriate bins. They then process the requests. Allocated requests are put into admitted bin (a list of allocated src-dst pairs) and the remaining traffic is then sent to the next core. The Allocator cores also keep track of time and appropriately send out the admitted bin and start working on the new timeslot. Cores performing so many different functions are not able to give good throughput.

%
%
%
%
%
%
%
%

\begin{figure}[h]
\centering
\includegraphics[scale=0.25]{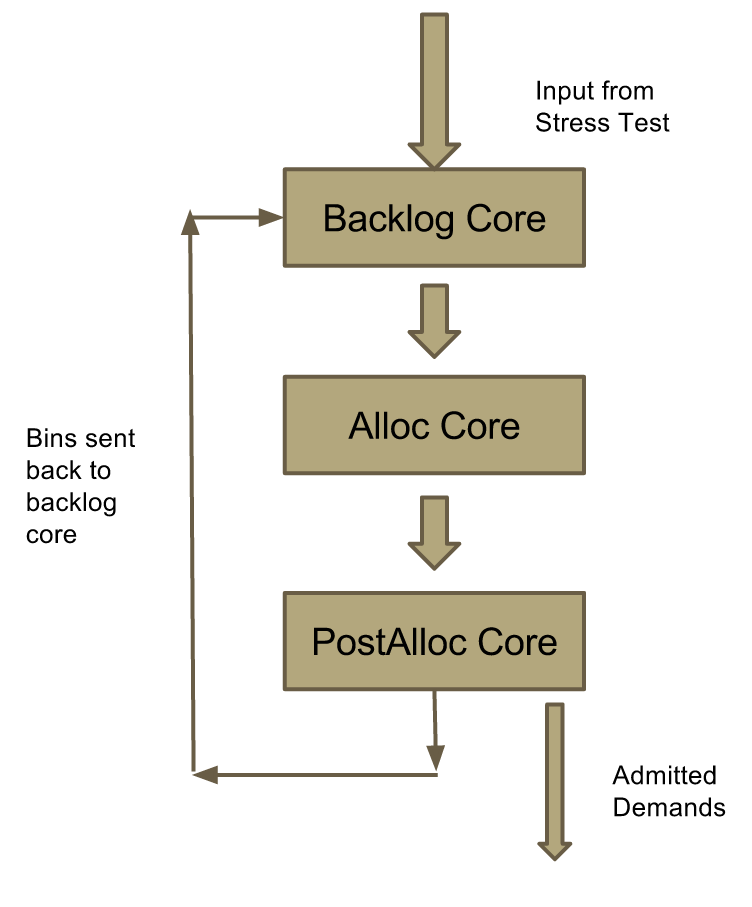}
\caption{Multiple cores sharing the load of a single alloc core}
\label{share}
\end{figure}

Instead of having a single core doing all the work, we share this load across multiple cores. Now each of the cores become simple and can sustain higher throughputs. We have a pipeline of cores working for the same timeslot as shown in figure ~\ref{share}. We separate one allocator core into three cores namely, the backlog core, the alloc core (new simple alloc core) and the post alloc core. We separate the backlog structure from the $CommCore$ and maintain it in the backlog core. The alloc core is now simple, it just gets a array of src-dst pairs, it needs to maintain the bit map and send out an array of bits denoting if the $i^{th}$ request is allocated. The post alloc core takes care of the job of splitting allocated demands from unallocated ones, sending back unallocated demands back to backlog core and send out the list of allocated src-dst pairs at the right time.  All these three cores are in a pipeline and they together perform allocations for a single timeslot. The communication between these cores is in the form of passing around bins. We fill requests into these structures called bins and we circulate these bins across these cores.

The backlog cores reads demand requests from the $CommCore$ (or the stress test core). It maintains a huge backlog structure which says if there is already an request from (src,dst) and if yes, it gives the number of timeslots that src-dst pair has requested. The backlog cores reads the incoming new demand requests and check against the backlog structure. If there is no outstanding request from that src-dst pair, then it adds the request to the bin. And the bin is sent out to the alloc core when its sufficiently filled. If the src-dst pair already has an request in the system, then it just increments the value in the backlog structure by the number of new timeslots requested (to avoid two requests for the same src-dst running around in the system). Once the old request (which is already in the system in one of the bins) get serviced fully, these backlogs are sent as new requests into the bins. 

The alloc core is written in $asm$ in C. It reads these request demands and fills a array of bits in the bin denoting if the $i^{th}$ request is allocated. The bin is then passed on to the post alloc core. Post alloc core separates the allocated, unallocated, spent and remaining demands and performs the other functions mentioned earlier. The bin is then sent back to the back log core which fills new demands. There are multiple bins being passed around so that the cores can work in parallel on different bins.

We no longer use DPDK queues for passing these bins between the cores. Instead, We use cachelines for communications. When the backlog core wants to send a bin to the alloc core, the backlog core writes the address of the bin to a cacheline and the alloc core reads from it. This is a single producer single consumer writing/reading from a single memory location problem which is solved by having a extra bit. The writer sets a bit after writing the new value. and the reader unsets the bit once done consuming the value. The writer writes only when the bit is unset and the reader reads only when it is set. The cachelines are used to transfer the pointers to the bins. Once the bin address is read, explicit prefetching optimizations were implemented to hide the latency of bringing the bin contents from the next core.

\begin{figure}[h]
\centering
\includegraphics[scale=0.25]{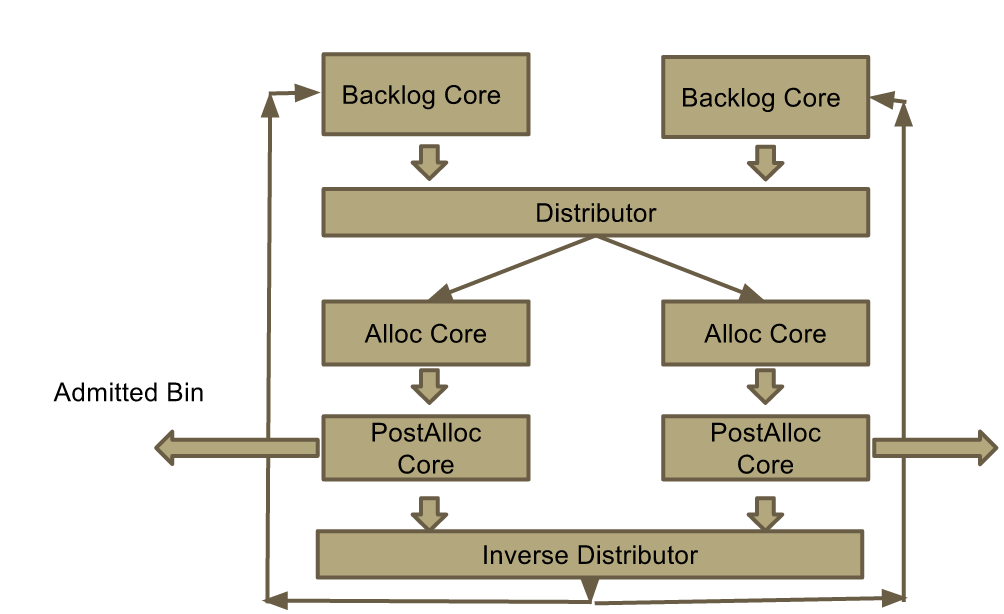}
\caption{Random Shuffle Architecture}
\label{shuffle}
\end{figure}

We discussed a way to make three cores work in a pipeline on a single timeslot. After this timeslot, they would all start working for the next timeslot. The actual scaling up is achieved when multiple such sets of three cores can be put together and made to work in parallel. Figure ~\ref{shuffle} explains the Random shuffle architecture which does this. Here we have multiple backlog cores each responsible for a subset of source nodes. In the example below, the first backlog core may be responsible for source nodes 1-127 and the second backlog core for 128-255. This means that any demand request from those sources go to the respective backlog cores and these backlog cores hold the backlog structures for those sources. This is very flexible design as when the number of nodes get huge, a single $CommCore$ cannot handle all the incoming requests and we may need multiple $CommCores$ each working on a subset of source nodes. In such a setting, each $CommCore$ can blindly send its requests to one particular backlog core that handles those source nodes.

The AllocCore1  and the PostAllocCore1 would be allocating for timeslot 1 and the AllocCore2 and PostAllocCore2 would be allocating for timeslot 2. If the bins from BacklogCore1 always goes to AllocCore1 and bins from BacklogCore2 always goes to AllocCore2, then fairness among the nodes would be lost. Any request coming from source nodes 128-255 always gets a latency delay of 1 timeslot more than the demand requests from source nodes 0-127 as they are processed only for one timeslot ahead in future. The problem would get worse when we have four such backlog cores as that would induce a three timeslot extra delay for some source nodes.

We introduce a distributor data structure between the backlog and the alloc cores that randomly permutes these bins from different back log cores across the different alloc cores. More importantly, the distributor provides a inverse distributor between the postalloc and the backlog cores, which guarantees that a bin originally sent from BacklogCore1 after getting processed at a random alloc core always reaches back to BacklogCore1 and never goes to any other backlog cores. The distributor and the inverse distributors implement prefetch optimizations so that the next incoming bin is already fetched at the core while it is processing the previous bin. Such optimizations hide the latency of transfer of bins between the cores.

%
%
%
%
%
%
%
%

\begin{figure}[h]
\centering
\includegraphics[scale=0.5]{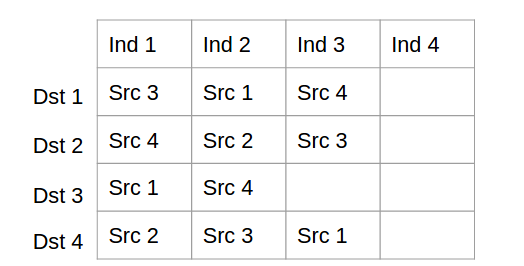}
\caption{Working of the distributor}
\label{distributor}
\end{figure}

The distributor is a two dimensional array of cachelines as shown in figure ~\ref{distributor}. Lets consider 4 sources (like BacklogCore1 to BacklogCore4) and lets call them Src1 to Src4. These sources want to randomly send bins to destination Dst1 to Dst4. Each of these sources has a bin to send. The first column in the figure denotes where the first bin from each of the sources go. For instance, the first bin from Src1 goes to Dst3, Src2 to Dst4, etc. and the second bin from Src1 to Dst1 and Src2 to Dst2. If we notice, every column of the table is a permutation of 1 to 4. For every $i^{th}$ bin from each of the Sources the distributor generates a random permutation and writes the bin address into the corresponding cacheline. The destination nodes read from these cachelines in order. Explicit Prefetching is done to bring the contents of the next bin into a core.

To generate the random permutation, we use a bit trick. Consider source/destination numbers from 0 to $2^n-1$. The $i^{th}$ bin from source number $x$ would go to the destination number $y$ where $y= A^(x+B) mod (2^n-1)$ where $A = (i * P1) mod (2^n-1) $ and $B = (i * P2) mod (2^n-1)$ and $P1, P2$ are known huge primes. This would give the random permutation that we required above.  The inverse distributor, would take $y$, $i$ and would get back $x$ by computing $x = (A^y)-B mod (2^n-1) $  where $A = (i * P1) mod (2^n-1) $ and $B = (i * P2) mod (2^n-1)$ and $P1, P2$ are the same known huge primes. This bit trick and the mathematical formulation gives us the required randomness and at the same time provides the flexibility to invert the permutation.

Benchmark
Within bin 
Across core

%
%
%
%
%
%
%
%

\subsection{Results}

To analyse the distributor and benchmark it, we wrote dummy backlog, alloc and postalloc cores where the alloc cores just dequeue bins from the inverse distributor and enqueue it into the distributors. The alloc cores works on the bins for a small amount of time (to make sure the advantages of prefetching is seen) and enqueues it off to the dummy postalloc core. The time latency for passing a bin around across these cores are calculated and reported in table \ref{res_dist}. This reduced latency of communication is one main backbone of our random shuffle architecture.
\begin{table}[!t]
\renewcommand{\arraystretch}{1.3}
\caption{Inter-core communication latency with the optimized distributor in micro seconds}
\label{res_dist}
\centering
\begin{tabular}{|c|c|}
\hline
Number of Cores & Latency for one bin circulation \\
\hline
 1 set of cores (3 cores) & 0.056 \\
\hline
2 set of cores (6 cores) & 0.188 \\
\hline
4 set of cores (12 cores) & 0.373 \\
\hline
8 set of cores (24 cores) & 0.920 \\
\hline
16 set of cores (48 cores) & 1.282 \\
\hline
\end{tabular}
\end{table}


With the super fast distributor in place, the Random Architecture was implemented and tested. a single stress test core wasn't able to generate the load for the entire system and hence, multiple stress test cores were put in place. Each of the stress test cores were responsible for a portion of the total number of nodes. Techniques were implemented to ensure within bin fairness. The system gave 2.15 Tbps on just one set of 3 cores as opposed to our baseline that gave 1.27 Tbps on 8 cores. The system scaled linearly while adding more sets of cores. It handled 7.1 Tbps on 4 sets of cores (12 cores) during our stress tests. The stress test environment were the same as discussed earlier.  From 8 cores to 16 cores, we do not get linear scaling. 16 Cores would span across 2 NUMA nodes on Ben and the inter-node-communication through QPI slows us down.


\begin{table}[!t]
\renewcommand{\arraystretch}{1.3}
\caption{Stress Test throughput for Random Shuffle Architecture in Tbps}
\label{res_shuffle}
\centering
\begin{tabular}{|c|c|}
\hline
Number of Cores & Throughput \\
\hline
 1 set of cores (3 cores) & 2.15 \\
\hline
2 set of cores (6 cores) & 4 \\
\hline
4 set of cores (12 cores) & 7.1 \\
\hline
\end{tabular}
\end{table}

\newpage
\section{Conclusions}
We worked on Fastpass, a data center network architecture aiming at zero queuing. The timeslot allocator of Fastpass was able to support only 1.27 Tbps of network traffic on 8 cores. We started off by analysing the existing pipelined architecture and suggested different changes to it including the batch processing mode that gave 1.5x better throughput. We then designed a parallel architecture which gave 1.5x better throughput with just half the number of cores. As the inter-core communication overheads in DPDK queues were high, we designed our own distributor data structure using cachelines to communicate between the cores. We further designed and implemented a Random Shuffle Architecture that supports up to 7.1 Tbps of network traffic allocations on just 12 cores.


\chapter{ECN BASED CONGESTION CONTROL FOR A SOFTWARE DEFINED NETWORK}
\label{chap:SDN}

\section{Abstract}

This chapter deals with congestion control in a software defined network
(SDN) setting. Presently, explicit router schemes, such as Explicit
Congestion Notification (ECN), work in conjunction with the TCP
protocol to handle congestion in a distributed manner.  With the
emergence of SDN and centralized control, it is possible to leverage
the global view of the network state to make better congestion control
decisions. In this work, we explore the advantages of bringing in
global information into distributed congestion control. We propose a
framework where the controller with its global view of the network
actively participates in the congestion control decisions of the end
TCP hosts, by setting the ECN bits of IP packets appropriately. Our
framework can be deployed very easily without any change to the end
node TCPs or the SDN switches. We also show 30x improvement over the
TCP Cubic variant and 1.7x improvement over TCP/RED in terms of flow
completion times for one implementation of this framework, using the
Mininet emulator.


%

\newpage
\section{Introduction}

This chapter deals with congestion control in computer networks.
Existing solutions can be categorized as end-node based or
router-based. The latter solutions use queue management and scheduling
algorithms that provide signals to the end hosts, to reduce the source
traffic. Active Queue Management techniques drop/mark packets at the
switch/router buffers thereby signalling the end nodes about
congestion [9, 10, 15]. There are some schemes that use both
active queue management and end node TCP modifications
[7, 11].


Most of the existing congestion control algorithms [4, 5] have a very limited view of the network and its
traffic. Many TCP based congestion control algorithms use packet loss
as an indicator of congestion.  Another measure used is the round-trip
time (RTT). The problem with using RTT is that, the feedback may be
easily misinterpreted. For, example consider a 100 packet backlog
(with 1,500 Byte packets) in a router queue. It corresponds to
$1,200~\mu s$ of queuing delay at 1~Gbps, but only $120~\mu s$ at
10~Gbps. The end node cannot make fine distinctions without more
information. Without the detailed knowledge of the underlying network,
TCP will continuously keep increasing and decreasing its congestion
window trying to adapt to the network, but may never end up doing so,
due to its parochial view.

TCP has been designed to work in a broad range of networks. Each TCP
variant works well for some kinds of the network and its traffic and
the same TCP performs poorly for other conditions. The interesting
part is that, we do not know exactly what objective does TCP
congestion control try to optimize [4]. This inflexibility
in adapting to new scenarios limits its use. 


The emergence of Software Defined Networking (SDN) gives network
protocol designers the power of centralized view and centralized
control that can be exploited for many applications
[14, 16].  SDN provides a centralized view of the network
with access to the statistics and other information of the routers and
link states. The central controller can aggregate these information
and actively participate in the congestion control decisions of the
end nodes.  The scope of this chapter is to study to what extent we can
exploit the central view and the centralized control features to
improve congestion control in networks.

An SDN-enabled scheme for handling congestion control is presented in
the thesis.  With a global view, the controller knows exactly what each
of the link states are and would never misinterpret a packet error as
congestion (as was done by many TCP variants). The information at the
controller can supplement the indicators like packet loss and delay,
that were used by the end nodes earlier. The controller can provide a
more realistic view of the network to the end nodes.  With a more
detailed knowledge about the network and the traffic flowing at any
point in time, we can take better, faster congestion control
decisions. In the proposed mechanism, the controller instructs the
switches (via the OpenFlow API) to set the relevant ECN bits on
packets going through a switch. This information is then used by the
TCP end-nodes for changing the TCP congestion window. The scheme has
been implemented in the Mininet emulator [17] and studied
for three different network scenarios. The results show that the
proposed approach achieves improvements over TCP CUBIC and TCP/RED
based distributed solutions.



\newpage

\section{Proposed SDN-based Framework}
%
%
%
%
%
%
\begin{figure*}[h]
\centering
\includegraphics[scale=0.25]{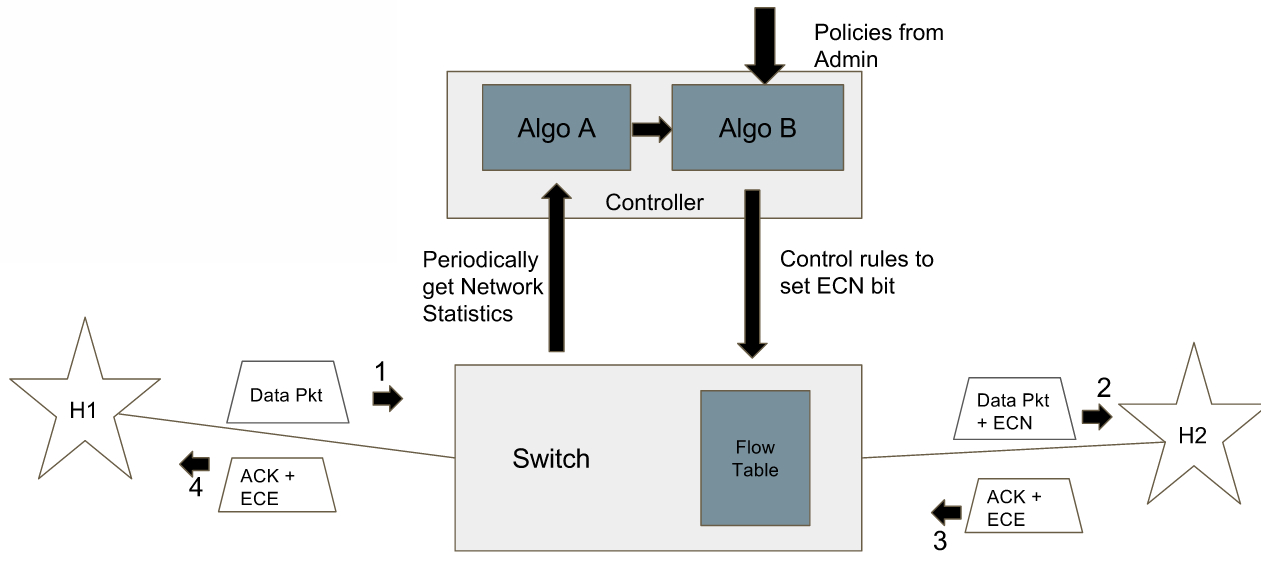}
\caption{Proposed SDN-based Congestion Control Framework.}
\label{fig_Arch}
\end{figure*}

The proposed SDN-enabled framework is shown in
Fig.~\ref{fig_Arch}. The network consists of several switches/routers
that implement the \textit{data plane} with the \textit{control plane}
implemented by a logically centralized SDN controller.  The end-nodes'
TCP protocol implementation is not modified, but is expected to be
ECN-enabled.  An SDN controller application periodically collects
information about the underlying network and the traffic
characteristics. Based on these information, it detects congestion
according to an algorithm (generically called {\bf Algorithm-A}).

With the global view of the network available, another algorithm
(called {\bf Algorithm-B}) selects the end nodes that need to react to
any detected congestion. This algorithm also considers the policies
and the priorities to flows set by the administrator to decide which
end nodes to penalize. After taking these decisions, the application
sends new flow rules to the switches (which timeout after a transient
amount of time), instructing them to set the ECN bits for particular
flows. The application conveys information regarding congestion to the
TCP end hosts by setting ECN bits. The normal TCP end nodes, react to
packets marked with ECN bits by reducing their congestion
window. Different implementations of \textit{Algorithm-A} and
\textit{Algorithm-B} can be used for different types of networks and
workloads.

\subsection{Congestion Detection} The congestion detection component
(\textit{Algorithm-A}) periodically collects information about the
network and its traffic and tries to recognize/predict congestion. SDN
switches have the capability of collecting statistics. The application
can probe the switch for these statistics and get a complete picture
about the network and its present traffic conditions. Queue statistics
and link utilization statistics will be used by \textit{Algorithm-A}
for detecting congestion. A simple metric to identify potential
congestion is if the link utilization or average queue length is
greater than a threshold. More sophisticated algorithms could consider
global network state and predict congestion based on learning
algorithms or other complex heuristics. In the evaluation section, we
will present an example for \textit{Algorithm-A}.

\subsection{Handling Congestion} 
Once \textit{Algorithm-A} predicts congestion in a link/switch,
\textit{Algorithm-B} decides on how to handle the congestion.  The
administrator can also set policies for congestion control. For
instance, the policy can specify that certain type of high-priority
flows should not be affected.  It will compute different ways to avoid
congestion, based on the global view and administrator policies. It
will identify the set of end nodes whose congestion window can be
reduced to avoid congestion while maximizing a global objective.

In traditional TCP, all the end nodes react to congestion by reducing
their congestion window. In data center networks and other SDN
applications, it is not necessary that all the flows should react to
congestion. For instance, we would like the interactive traffic or
short-term (i.e. \textit{mice}) flows to be not modified but control
the congestion windows of bulk (i.e. \textit{elephant}) traffic. More
over, when all the nodes react to congestion, they all back off at the
same time (leading to under utilization) and then move forward
simultaneous giving rise to a saw-tooth like bursty traffic at the
switches. 

For example, the following scheme can be used as \textit{Algorithm-B}
in data center networks for prioritizing interactive flows.  Whenever
congestion is detected in a link, the framework will obtain all
statistics about the flows traversing that link. The algorithm can
select the top $T\%$ of the the flows based on their bandwidth
utilization and reduce their congestion windows. Generally, only about
10\% of the flows are elephant flows in the network, but they utilize
about 90\% of the bandwidth. Penalizing the high utilization flows can
ensure that interactive traffic is less affected by congestion.

\subsection{Congestion Notification}

The next step is to convey the congestion handling information to the
end nodes so that they can react appropriately. The
\textit{Algorithm-B}, after computing the set nodes that have to
reduce their congestion window, will send explicit flow rules to the
switches with an additional action of marking the ECN bits apart from
forwarding the packets to the right port.  These new high priority
rules sent by \textit{Algorithm-B} are called the Congestion Control
Flow Rules (CC flow rules). When new packets from these flows arrive,
they get matched with these top priority CC flow rules and their ECN
bits get set at the switch before forwarding. 

The ECN bit processing is based on the end-nodes' TCP
implementation. As mentioned earlier, we do not require any changes in
the TCP implementation.  A receiver, upon receiving packets with
marked ECN bits, echoes the information to the sender by marking the
ECE bit in their ACK packet. The sender acts upon these marked ACKs
and reduce their congestion window according to their native TCP
variant.  

The new ECN marking flows added are given higher priority in the
OpenFlow tables at the switches, compared to the regular flows. This
is to make sure they are not skipped. Also, these new CC flow entries
are set with a very small rule timeout interval so that adequate
number of packets are marked. If a new CC flow rule is added for a
very long time, it might send signals of large congestion to the end
node that would affect the throughput of that flow adversely. At the
same time, the timeout interval should not be too low such that not
enough packets from that flow actually get marked. Choosing this
timeout value is a critical parameter. Once the timeout interval is
over, these extra high priority flows get evicted and the flows start
matching with the normal low priority flows rules that were already
present in the switches.

Another critical parameter is the periodicity of the network probes.
As mentioned earlier, \textit{Algorithm-A} probes the network for
network and traffic state. There should be enough time for the system
to settle down after the congestion window changes are implemented by
the source TCP and the network traffic stabilizes. The ECN packets
have to reach the receiver and the receiver has to reply back with
bits marked in the ACK packet. It clearly takes about 2 RTT to get
communicated to the end nodes and it might take a couple of more RTTs
for the network to become stable. Very large probing interval cannot
detect sudden congestion and can be less efficient. Thus, the probing
interval is a sensitive parameter.

In summary, the salient features and advantages of this framework
include:

\begin{enumerate}
\item {\bf Global View:} This enables the congestion control scheme to
  obtain a more complete picture about the network state and hence
  make better congestion predictions that can achieve more globalized
  objectives.  In conventional end-to-end systems, all the end nodes
  try to optimize their own local objective function and could end up
  in a Nash equilibrium solution.

\item {\bf Prioritization in Congestion control:} Flow priority can be
  considered in making congestion control decisions and in setting of
  flow table rules.

\item {\bf Fairness:} Fairness can be ensured even across TCP and
  UDP. Typically, UDP flows can hog the bandwidth placing TCP flows at
  disadvantage. With the proposed approach, we can monitor the
  bandwidth used by UDP and use special flow rules to restrict UDP's
  uneven share of the bandwidth.

\item {\bf No change to the end-nodes:} The proposed approach does not
  require changes to the end-nodes' protocol stack.

\item {\bf No change to switches:} This approach uses normal flow
  rules and the functionalities available with existing SDN switches
  to achieve congestion control.

\item {\bf Easily pluggable CC algorithm}: The Algorithms $A$ and $B$
  can be changed according to different network and traffic needs. A
  data center network may need a different kind of algorithm than an
  enterprise network. Congestion control algorithm at the controller
  becomes a easily changeable mechanism.

\end{enumerate}

\newpage

\section{Performance Evaluation}

This section presents the performance evaluation of the proposed
congestion control framework and compared to existing schemes.

\subsection{Implementation Details}

The proposed framework has been implemented in an SDN emulator package
called Mininet (version 2.2.1) [17]. The Floodlight SDN
controller [18], Open vSwitch (version 2.3), OpenFlow 1.3
that supports setting of ECN bits through flow rules have been used.
The congestion control framework has been implemented as an
application in Floodlight. In all our experiments, the switch contains
a single flow table. All the virtual end hosts created by Mininet in
our experiments run TCP Cubic implementation that comes with Ubuntu
14.04 kernel. The ECN mechanism has been turned on at all the end
nodes. Since Mininet is not able to handle high bandwidths accurately,
100Mbps links have been used everywhere unless stated otherwise.

The analysis is done for data center networks with the objective of
achieving lower flow completion times (FCT) for interactive traffic
when it co-exits with bulk traffic.  As a proof of concept for our
framework, we present sample congestion control algorithms,
(\textit{Algorithm-A} and \textit{Algorithm-B}), and evaluate their
performance.

The proposed \textit{Algorithm-A} probes the switches every 2 seconds
and collects port statistics information. From these, we compute the
average link utilization of every link in that interval. If the link
utilization is greater than 75\% , we consider that congestion is
likely to occur in that link and inform \textit{Algorithm-B} about
it. The handling algorithm, \textit{Algorithm-B}, is defined as
follows. Congestion control is imposed on the top $T$\% of flows
(based on bandwidth utilization) going through the congested link once
the link utilization is greater than 75\%. We linearly increase $T$ as
link utilization increases and when link utilization reaches 100\%, we
set $T = 50\%$ penalizing top 50\% of the flows when utilization hits
the maximum. Since we are not penalizing all the flows at the same
time, this can prevent saw tooth like behavior of
utilization. Additionally, we start penalizing more and more as we get
closer to 100\% utilization making sure we have enough bandwidth for
new incoming short flows.


The proposed approach is compared with Linux's TCP Cubic scheme, and
approaches that use in-network elements in congestion control, namely
RED [10] and ECN [3]. The implementation of RED
and ECN is available in Mininet. 

\subsection{Throughput with long flows} The objective of this
experiment is to show that the proposed algorithm does not compromise
on bulk flows and achieves good throughput on long-lived traffic.

The topology studied, denoted \emph{Topology~1}, consists of one switch
connected to four end nodes with 100 Mbps links each. Three nodes are
senders ($S_1,S_2,S_3$) and one is a receiver ($R_1$).  Three of the
senders run \emph{iperf} for two minutes to generate traffic to the
receiver. The results are shown in Table~\ref{throughput_bulk}. As
seen, the proposed approach gives 8\% better total throughput than
Cubic and comparable performance with the distributed ECN and RED
schemes. The proposed method achieve better fairness than TCP Cubic
and is on par with ECN and RED.

\begin{table}[!t]
\renewcommand{\arraystretch}{1.3}
\caption{Throughput of bulk flows for Topology 1 (Mbps)}
\label{throughput_bulk}
\centering
\begin{tabular}{|c|c|c|c|c|}
\hline
Flow & TCP Cubic & ECN & RED  & Proposed \\
\hline
S1 &32.5 &29.6 &30.9 &32.7 \\
\hline
S2 &25.8 &29.4&28.1&29.7 \\
\hline
S3 &26.0 &30.2&32.2&28.7 \\
\hline
Total & 84.3&89.2&91.2&91.1 \\
\hline
\end{tabular}
\end{table}

The next topology, called \emph{Topology~2}, is a dumbbell topology
with 2 switches and 3 nodes connected to each switch. Nodes connected
to one of these switches are all senders, called $S_1, S_2, S_3$. The
nodes connected to the other switch are receivers $R_1,R_2,R_3$.
Three pairs ($S_1,R_1$),($S_2,R_2$),($S_3,R_3$) are selected and
long-lived flows are established between them for 2 minutes. All the
links have 100 Mbps capacity. The throughput results are shown in
Table~\ref{throughput_interactive}. As seen, the proposed approach
achieves 14\% better total throughput than TCP Cubic and is similar to
that achieved by RED and ECN. The individual through-puts are also
closer to the fair-share values.

\begin{table}[!t]
\renewcommand{\arraystretch}{1.3}
\caption{Throughput of bulk flows for Topology 2 (Mbps)}
\label{throughput_interactive}
\centering
\begin{tabular}{|c|c|c|c|c|}
\hline
Flow & TCP Cubic & ECN & RED &  Proposed \\
\hline
S1 &27.5&31.3& 32.9&31.3\\
\hline
S2 &33.3&29.9& 30.9& 31.9\\
\hline
S3 &23.3&32.9& 32.1&32.5\\
\hline
Total (Mbps)&84.1&94.1& 95.9&95.7 \\
\hline
\end{tabular}
\end{table}

\subsection{Coexistence of interactive and bulk traffic} In this
set of experiments, we show that the proposed algorithm significantly
improves the flow completion time of interactive flows in presence of
bulk traffic. In the first experiment, we consider the single
switch-four node \emph{Topology~1}.  Two long-lived flow between from
$S_1$ and $S_2$ to the receiver $R_1$ are established. A 2~MB
interactive flow is sent from $S_3$ to $R_1$. This experiment is
repeated 30 times for each of the congestion control algorithm and the
interactive flow's mean flow completion time is computed, as shown
below.

\begin{center}
\begin{tabular}{|c|c|c|c|c|c|}
\hline
TCP Cubic & ECN & RED &  Proposed \\
\hline
10.64 & 0.36 & 0.41 &  0.34 \\
\hline
\end{tabular}
\end{center}

As seen, the proposed scheme obtains 30x improvement to Cubic in flow
completion time and about 1.2x improvement to RED. ECN was close (but
worse) to our scheme in this topology. We also found that our
algorithm shows the least variance in flow completion time compared to
the other schemes, based on the 30 iterations.

The \emph{Topology~2}, discussed in the previous section, was next
considered and flows established as above. The flow computation time
results based on 30 iterations is shown below.

\begin{center}
\begin{tabular}{|c|c|c|c|c|c|}
\hline
TCP Cubic & ECN & RED &  Proposed \\
\hline
 10.41 &0.46 & 0.37 &  0.37\\
\hline
\end{tabular}
\end{center}

As seen, the proposed approach performs 28x better than TCP Cubic,
1.25x better than ECN, but on par with RED. It was observed that the
proposed schemes produce had the least variance among the four.

To summarize, the proposed scheme outperforms TCP Cubic by a large
factor. It works the best for both the scenarios, while ECN and RED
failed to perform better than the proposed scheme in at least one of
the cases.

%
%
%
%
%
%
\begin{figure}[!t]
\centering
\includegraphics[scale=0.35]{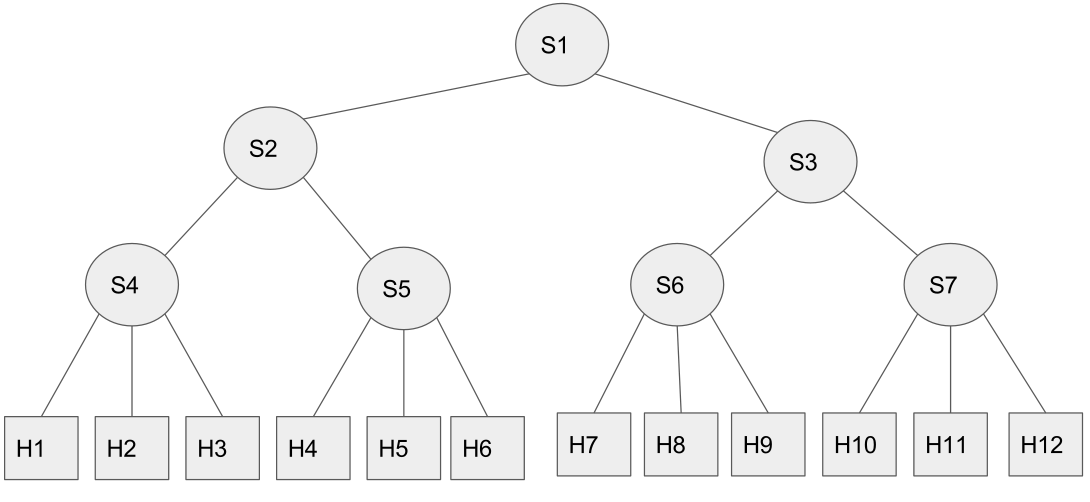}
\caption{Multi-hop Topology}
\label{fig_top}
\end{figure}

\subsection{Multi-hop scenario} 

We evaluate the proposed algorithm in multi-hop networks with multiple
bottleneck links. The 7-switch topology is shown in
Figure~\ref{fig_top}.  In the first experiment, we establish long
lived flows from $H_1,H_2,H_4,H_5$ to $H_7,H_8,H_{10},H_{11}$
respectively for 2 minutes. The total throughput (in Mbps) is
presented below. As seen, the proposed scheme achieves throughput
better than TCP Cubic and comparable to ECN, but less than that of
RED. 

\begin{center}
\begin{tabular}{|c|c|c|c|c|c|}
\hline
 & TCP Cubic & ECN & RED &  Proposed \\
\hline
Throughput & 76.7 & 83.9  & 85.4 & 83.8 \\
\hline
\end{tabular}
\end{center}
 
To check flow completion time for mice flows in this setting, We
created long-lived flows from $H_1,H_2,H_4,H_5$ to
$H_7,H_8,H_{10},H_{11}$ respectively. We simultaneously sent 2MB data
from $H_3$ to $H_7$ and from $H_6$ to $H_8$ and measured the flow
completion times. This experiment was repeated 30 times. The average
flow completion time (in seconds) achieved by each of the algorithms
are shown below. 

\begin{figure}[h]
\centering
\includegraphics[scale=0.5]{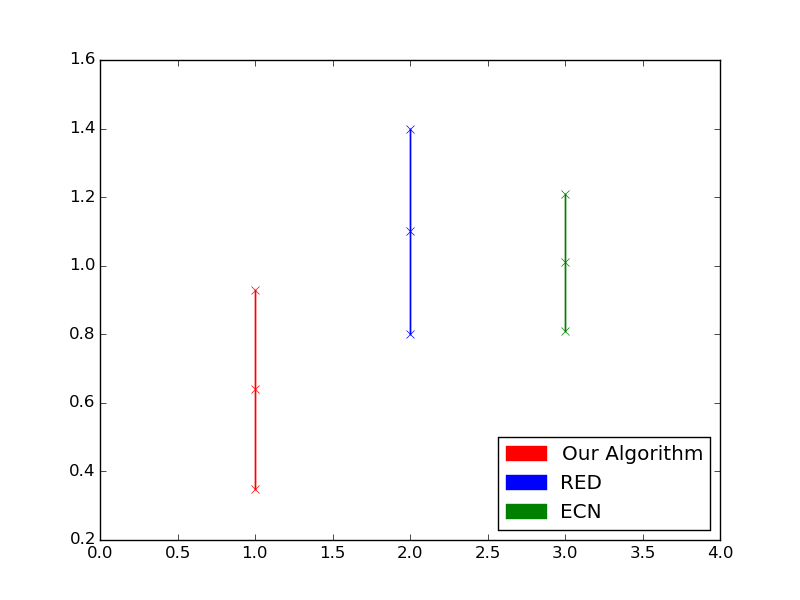}
\caption{90\% confidence interval plot for 30 trials of FCT in multi hop scenario}
\label{plot}
\end{figure}

\begin{center}
\begin{tabular}{|c|c|c|c|c|c|}
\hline
TCP Cubic & ECN & RED & Proposed\\
\hline
19.46 & 1.01 & 1.1 & 0.64 \\
\hline
\end{tabular}
\end{center}

In multi-hop conditions, our scheme clearly outperforms ECN by a
factor of 1.5x and RED by 1.7x without compromising much on throughput
of bulk flows. Our scheme also outperforms TCP Cubic by 30x.







\newpage
\section{Conclusions}

This chapter has presented a simple and easy to deploy framework for
congestion control in Software Defined Networks. The framework is
extensible in terms of specific congestion control and detection
algorithms. It requires no changes to the end nodes or the SDN-enabled
OpenFlow switches. The proposed framework has been implemented in the
Mininet emulator, with heuristics for data center networks. The
proposed approach shows 30x improvement to TCP Cubic, 1.7x to RED and
1.5x to ECN on multi-hop topologies for flow completion times of
interactive traffic without compromising throughput of bulk flows.

\chapter{CONCLUSION}

We saw that data centers networks require properties of: high utilization, low median,tail latencies and fairness. Specifically in data centers, the bulk and interactive traffic co-exist. Prioritizing interactive traffic over bulk traffic becomes essential. Owing to the smaller scale of operation of datacenters as compared to the internet, the concept of central control can be exploited. 

We first dealt with Fastpass, a data center Internet architecture. Even though it enables us to prioritize traffic with a central arbiter, it was not able to scale beyond 1.27 Tbps on 8 CPU cores. We analysed its existing pipelined allocator architecture and suggested changes enabling 1.5x better throughput. We also designed and compared the parallel architecture and the Random Shuffle architecture. We finally were able to scale the arbiter linearly till 12 cores supporting 7.1 Tbps of network traffic allocations. 

In the second part of the thesis, we dealt with the problem of congestion control in a Software Defined Network. We proposed a ECN based congestion control framework for SDN setting which requires no change to the SDN switches or the end nodes. We also showed 30x improvement over TCP cubic and 1.7x improvement over RED in flow completion times of interactive traffic for one implementation of this framework.









\end{document}